\documentclass[manuscript]{aastex}

\usepackage{natbib}

\shorttitle{Something will be here}
\shortauthors{Bertello et al.}

\begin{document}


\title{Signature of Differential Rotation in \\
    Sun-as-a-Star Ca II K Measurements}


\author{L. Bertello}
\affil{National Solar Observatory, Tucson, AZ 85719}
\email{bertello@nso.edu}

\author{A. A. Pevtsov}
\affil{National Solar Observatory, Sunspot, NM 88349}
\email{pevtsov@noao.edu}

\and

\author{A. Pietarila}
\affil{National Solar Observatory, Tucson, AZ 85719}
\email{apietarila@nso.edu}


\begin{abstract}
The characterization of solar surface differential rotation (SDR) from disk-integrated
chromospheric measurements has important implications for the study
of differential rotation and dynamo processes 
in other stars. Some chromospheric lines, such as Ca II K, are very sensitive to
the presence of activity on the disk and are an ideal choice for investigating SDR in Sun-as-a star
observations. Past studies indicate that when the activity is low, the determination of Sun's
differential rotation from integrated-sunlight measurements
becomes uncertain. However, our study shows that using the proper technique,
SDR can be detected from these type of measurements even during periods of extended solar minima.
This paper describes results from the analysis of
the temporal variations of Ca II K line profiles observed by the Integrated Sunlight Spectrometer (ISS)
during the declining phase of Cycle 23 and
the rising phase of Cycle 24, and  
discusses the signature of SDR in the power spectra
computed from time series of parameters derived from these profiles. The described methodology is quite general,
and could be applied to photometric time series of other Main-Sequence stars 
for detecting differential rotation.

\end{abstract}



\section{Introduction}

Rotation plays a fundamental role in processes of stellar formation and evolution,
through the modification of hydrostatic balance and the redistribution of
chemical elements (see Maeder \& Eenens 2004 for a review of stellar rotation).
%%chemical elements (see \citet{2004IAUS..215.....M} for a review of stellar rotation).
The Sun is the only star for which surface differential rotation can be studied in great detail,
by using spectroscopic Doppler-velocity measurements or by tracking the motion of
features such as sunspots (Beck 2000). Furthermore, the internal rotation of the Sun can be probed extensively 
using the techniques developed through helioseismology, that is, the analysis of observed oscillations
of the solar surface (e.g., Thompson et al. 2003).
Although solar surface differential rotation (SDR) is still not fully understood, it is
thought to be a key ingredient of the global-scale solar dynamo that generates the 22-year cycle of magnetic
activity (Schrijver \& Zwaan 2000).

Stellar differential rotation 
cannot be observed directly since the surface of stars usually cannot be imaged with sufficient
resolution. However, one may expect active stars to be rotating differentially
assuming that Sun-like dynamos also work on other stars.
Several photometric and spectroscopic techniques have been developed to investigate the rotation of distant stars. 
Photometric methods are designed to detect the signature of spots in stellar light-curves. If a 
star is rotating differentially, large 
starspots located at different latitudes may produce modulations in the light-curve signal that can lead
to the detection of differential rotation (e.g., Fr\"ohlich et al. 2012; Frasca et al. 2011). 
Flux measurements in the calcium H and K spectral lines have
also been successfully employed to monitor chromospheric stellar activity and detect rotation (e.g., Donahue et al. 1996).
Spectroscopic techniques measure the change in wavelength position and line profile of rotationally-broadened absorption
lines produced by the spot-observer relative motion due to stellar rotation (e.g., Barnes et al. 2005; Ammler-von Eiff
\& Reiners 2012). The deployment of asteroseismic space projects, such as the NASA Kepler mission (Koch et al. 2010), 
make
the detection of stellar differential rotation even more accessible.
The Kepler satellite provides high-quality optical photometric light-curves of unprecedented precision and duration (Borucki et al. 2010; Koch et al. 2010) that routinely allow detailed studies of stellar magnetic activity on late-type stars.
Rotational modulation due to starspots is commonly seen in the Kepler light-curves of these stars, 
allowing detailed study of the surface distribution of their photospheric magnetic activity
(e.g., Fr\"ohlich et al. 2012; Frasca et al. 2011; Savanov \& Dmitrienko 2012; Walkowicz \& Basri 2011). 
The multi-year duration light-curves provided by the Kepler mission will make it possible to investigate 
in detail how activity phenomena -- such as starspot evolution, differential rotation, and activity cycle --
operate on stars with a wide range of masses and convection zone depths.

Assessing the properties of SDR from disk-integrated measurements has
a very important diagnostic value for the interpretation of stellar observations. 
Since the Sun can be observed in both disk-integrated and disk-resolved modes, solar observations can 
serve as a test-bed for different techniques that can be applied to other stars in search for their 
differential rotation. Here we 
present a very effective method for extracting the rotational components from time
series of photometric measurements. We have applied this method to solar disk-integrated chromospheric
and photospheric time series, allowing us to compare the detected rotational periods to
the well known profile of SDR. 

Different solar activity
indices and methods have been used in previous studies in order to unambiguously extract the signature 
of SDR from Sun-as-a-star data. 
Hasler et al. (2002) used UARS SOLSTICE data and a combination of wavelet and Fourier
analyses, but beside the basic period of $\sim$27 days their results were inconclusive.
Hempelmann \& Donahue (1997) applied wavelet analysis to disk-integrated Ca II K line core emission
time series covering the duration of Cycle 21 and 22. Their results match the SDR profile only qualitatively.
In a later study, Hempelmann (2002) used a similar time-frequency analysis on SOLSTICE UARS data and found a clear
signature of SDR from the end of Cycle 22 until the beginning of Cycle 23. However, SDR was not visible in the
foregoing part of Cycle 22. Attempts have been made using white-light measurements, but the results are
negative (Hempelmann 2002).
Donahue \& Keil (1995) reported a successful detection of SDR from the analysis of
chromospheric full-disk Ca II K observations made almost daily, since 1977, by the National Solar Observatory at 
Sac Peak K-line  monitoring program (later referred to as SPO data, 
Keil \& Worden 1984). 
Donahue \& Keil divided the time series in 200-day intervals and computed
the power spectrum of the individual segments
using the Lomb-Scargle periodogram. Figure 3 of Donahue \& Keil (1995) shows the observed solar period
during Cycles
21 and 22. However, this result was criticized by Schrijver (1996) since the derived sidereal rotation period
of 28.5 days at the beginning of Cycle 22 is too large and the observed decline in rotation period during that
cycle is too steep. 
An additional limitation in the analysis is that the Lomb-Scargle periodogram typically
produces a large number of peaks spread over a broad spectral band. In their investigation, however,
only the periodicity corresponding to the location of the peak with maximum power within a small spectral window
centered at around 27 days was considered. Peaks outside that frequency range were ignored.

Ideally, the power spectrum of time series containing the possible signature of SDR
should have a minimum number of peaks to facilitate the interpretation of the different
components of the signal. This is particularly important when prior knowledge of the actual rotation period
is not available, e.g., in the case of stellar differential rotation. 
Our choice of spectral estimator was made to specifically address this point.
The detection of stellar rotation hinges on the presence of active features (e.g., starspots), and
properties such as lifetime and emergence rate of these features may play a critical role in the interpretation
of the observed signal. 
In the next section we describe the main properties of the time series investigated in this study. In Section
3 we delineate the main steps of our data analysis, with particular emphasis on the choice of spectral
estimator and computation of statistical significance of the derived estimates. The results of the
data analysis are discussed
in Section 4,
which also includes the description of a simple model to estimate 
the effects of magnetic field diffusion and the rate of sunspot emergence on the
detection of solar rotation in Sun-as-a-star observations.

\section{The ISS Ca II K parameter time series}

In this study we use observations from the Integrated Sunlight Spectrometer (ISS), one of three instruments comprising 
the Synoptic Optical Long-term Investigations of the Sun (SOLIS, Balasubramaniam \& Pevtsov 2011 and reference therein). The ISS, in operation
since late 2006, takes high spectral resolution (R $\cong$ 300,000) daily observations of the Sun-as-a-star in nine different wavelength bands.  For this study we select the observations taken in the Ca II K spectral line centered at 393.37 nm, the longest continuous data in the ISS data set. 
A least-square fit to known wavelength positions of five photospheric spectral lines situated in the red and blue wings of the Ca II K line is performed to determine the linear dispersion. The error in determination of dispersion is extremely small to the point that one can see slight ~0.02\% annual variations that are attributed to 
seasonal changes in the refraction index of air (Bertello et al. 2011).
In late 2011, the ISS was upgraded to include optical encoders on the spectral grating and mechanisms moving the CCD camera (for flatfielding purposes). This upgrade further improved the ISS stability. 

The spectral band for the Ca II K line covers about 0.05 nm, and thus, does not include the continuum near the Ca K line. To overcome this limitation, the observed line profiles are normalized  using intensities at two narrow bands situated in the blue (393.147-393.153 nm) and red (393.480-393.500 nm) wing of the Ca K line. Mean intensities in these two bands are scaled to match intensities in the "reference" spectral  line profile taken by the NSO Fourier Transform Spectrometer (FTS, Wallace et al. 2007). This normalization also helps to remove any  residual linear gradients in intensity in the spectral direction. The normalization bands are situated far in the wings of the K line, and thus, are not likely affected by solar cycle variation. As a test, we examined intensities in the cores of photospheric lines situated near the bands used for scaling the intensities, and did not find any systematic trends indicating a solar cycle dependency. For a  more detailed description of the ISS data reduction we refer the reader to Bertello et al. (2011).

From each spectrum the nine parameters listed in Table \ref{tbl} are extracted. These parameters 
may have contributions from different heights and/or solar features, and
their response to solar activity is not
identical. As described in Bertello et al. (2011),
K3 intensity, 0.5-\AA~ and 1-\AA~ emission indices are highly correlated with each other.
Wavelength separation of V \& R emission maximum does not correlate well with any other parameters, 
while K3 wavelength shows moderate correlation with most of the other parameters.

\section{Data Processing}

From December 2, 2006 to January 27, 2012 the ISS instrument has taken
2078 measurements in the Ca II K spectral band. Each individual
spectrum is normalized according to the procedure described in the previous
section and the set of nine parameters listed in Table \ref{tbl} is computed. 
Time series of these parameters are updated as new daily observations are taken 
and can be accessed from the SOLIS-ISS web site at
http://solis.nso.edu/iss.

\subsection{Data Rejection}

\begin{figure}
\plotone{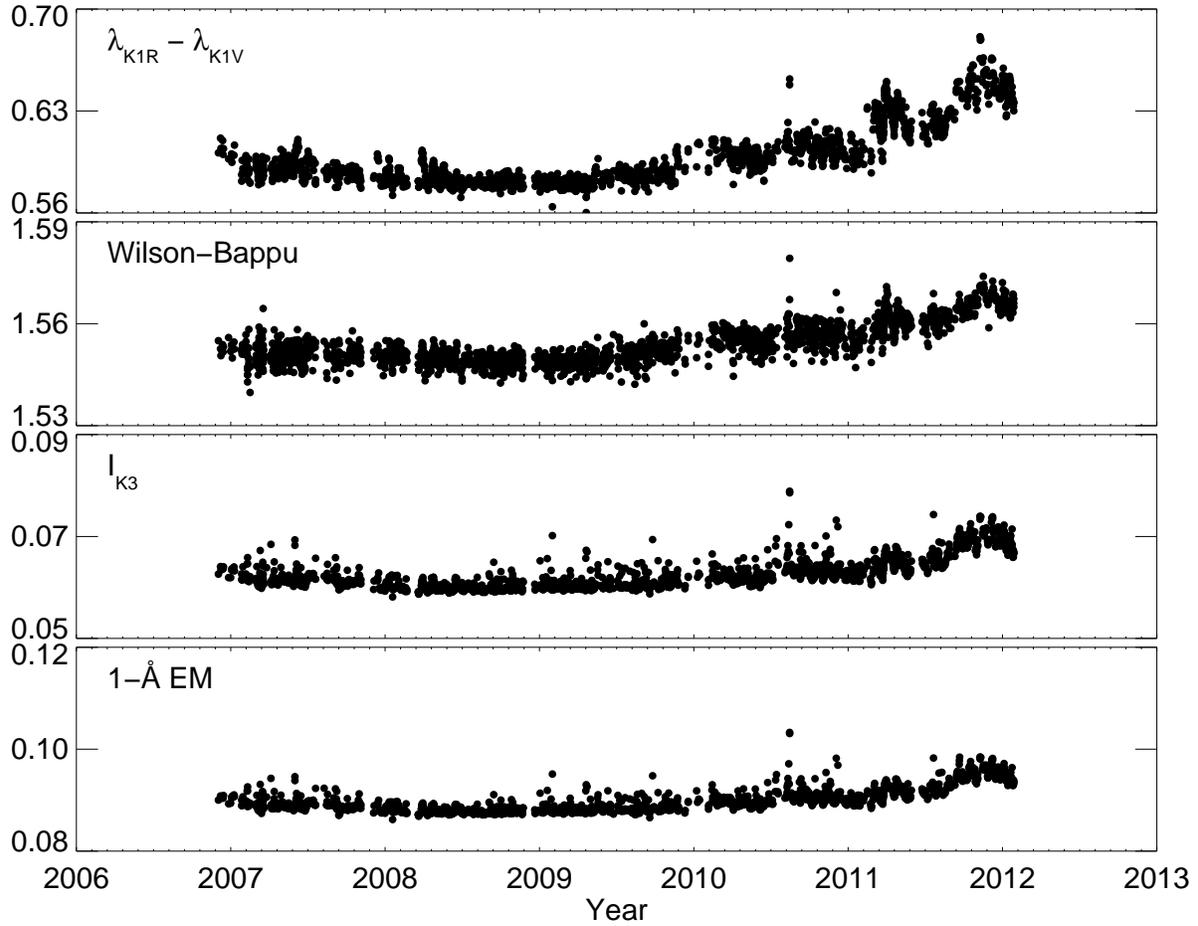}
\caption{Four of the eight ISS Ca II K parameter time series listed in Table 1.
The period of time covered by these
observations is from December 2, 2006 to January 27, 2012.
Daily updates of all parameter time series are available from the ISS web site
at http://solis.nso.edu/iss.
\label{para_ts}}
\end{figure}

Figure \ref{para_ts} shows the temporal variation of
some of the parameters investigated in this study. The other parameters exhibit a similar behavior,
with the exception of $\lambda_{\rm K3}$ that does not show any significant temporal trend and was
therefore excluded from our analysis.
The rising phase of Cycle 24 is clearly visible in the
temporal behavior of these parameters. 
The error bars of the individual
measurements are typically smaller than the size of the points shown in the figure,
and are omitted for clarity. 
Although the error bars for individual measurements are small, there 
is a small number of measurements which deviate significantly from the general population. 
These outliers were traced to specific (bad) observations. 
The first step in the data reduction consists of rejecting
the outliers from the time series. This was achieved by fitting a cosine function with period of
$\sim$11 years to the data and
eliminating points that are 3-$\sigma$ away from the model. The
procedure was repeated until no points were eliminated. The number of points rejected 
depends on the particular time series, varying from 14 (I$_{\rm K2V}$-I$_{\rm K3}$)/(I$_{\rm K2R}$-I$_{\rm K3}$)
to 44 (I$_{\rm K2V}$/I$_{\rm K3}$) out of 2078 total points. 

\subsection{Resampling and Filtering}

Each clear day the ISS takes typically 1-2 observations in the Ca II K bands,
producing time series of parameters that are not evenly sampled in time. 
The problem of estimating the power spectral density from irregularly sampled data is in general
more complicated than that of equidistant data. The direct Fourier transform and the method
of Lomb-Scargle converge to the usual periodogram if the observations are equidistant, but suffer
from a significant bias in the estimated spectra of unevenly sampled time series (Benedict et al. 2000).
Furthermore, the Lomb-Scargle periodogram (Scargle, 1982) is particularly designed to detect sinusoidal
signals in white-noise unevenly time series.
However, observational data such as those investigated here often contain fractions of non-Gaussian noise or
may consist of periodic signals with non-sinusoidal shapes. These properties of the data
make the interpretation of the Lomb-Scargle periodogram more difficult and can lead
to misleading estimates. Depending on the properties of the time series, many resampling methods 
are available (e.g., de Waele \& Broersen 2000) that allow the use of more effective spectral estimators.
Nearest neighbor resampling with the slotting principle (Broersen, 2009) is a technique 
that replaces an irregularly spaced sample by an equidistant signal with the resampling distance
$\Delta$. At every resampling node, the closest irregular observation is substituted if it is
within half the slot width $\Delta$ from the resampling time. If no irregular observation falls within 
the slot width, the grid node is left empty as a missing observation. If the duty cycle of the
resampled time series is reasonably high, missing observations can be filled via linear interpolation
without introducing spurious peaks in the spectral estimator (de Waele \& Broersen 2000). Because
of the high temporal coverage of the ISS observations, a suitable
value for the slot width is $\Delta$ = 1 day, with the resampling time set at Noon local time. 
This particular
choice for the resampling time selects the daily ISS observation with the best signal-to-noise
ratio.

To eliminate spectral leakage from the very low frequency range, each time series was detrended
using a 4-th order B-spline. Next, the signals were digitally filtered by means of a low-pass
finite impulse response 
linear phase filter designed according to the recipe described in McClellan et al. (1979). The application
of the filter before the spectral estimate is necessary to ensure that signal distortions due to aliasing
are negligible. The filter
was designed to have a passband between 0 and 0.2 cycles/day (5 days), a transition band between
0.2 and 0.4 cycles/day (2.5 days), and a stopband above 0.4 cycles/day.
This was achieved using 21 coefficients, which gives an attenuation of $\sim$65 db for the
stopband and a maximum deviation of $\sim$0.001 db in the passband. 
Finally, the data were pre-whitened to reduce the non-stationary components
of the signal.

\subsection{Spectral Estimation \label{spect}}

Among the available spectral estimators we selected 
the maximum entropy spectral estimator (MEM), which is equivalent to the autoregressive (AR)
spectral estimator (Brockwell \& Davis 1991). The maximum entropy spectral
estimator is well suited for the analysis of our data because of its high spectral resolution, even with relatively
short time series. If $\{x_i\}$ is a zero-mean Gaussian stochastic process $N(0,\sigma^2)$, which 
models our time series, the spectral density $S(\nu)$ at frequency $\nu$ is given by

\begin{equation} 
S(\nu) = \frac{\sigma^2}{|1 + \sum_{k=1}^{p}a_k\exp(-ik\nu)|^2},
\label{mem}
\end{equation} 

where $a_k$ and $p$ are the coefficients and order of the AR process:

\begin{equation}
x_i = \sum_{k=1}^{p} a_k x_{i-k} + z_i.
\label{arcoef}
\end{equation} 

Different estimation procedures exist as to how the AR coefficients, $a_k$, and the
white noise variance, $z_i$, are estimated in Equation \ref{arcoef}
for any order $p$ (e.g., Robinson \& Treitel 2000, Brockwell \& Davis 1991, Percival \& Walden 1993).
The optimal choice of $p$ needs to be considered carefully: if the number of poles, $p$, is too
high, the method can introduce spurious and/or split peaks when applied to a noisy time series
with a very large number of data points. A number of criteria have been suggested for selecting
$p$ (e.g., Brockwell \& Davis 1991), but in most applications $p$ is chosen empirically
(Papoulis 1991). For our study we verified that as long as $p$ is inside a reasonable
interval, $p \in [N/4,N/2]$ ($N$ being the number of data points), the same spectral features
are found to be statistically significant for all values of $p$. A value of $p = N/3$ was used in
our analysis.

Unlike more traditional approaches based on Fourier techniques, the MEM lacks an easy and 
self-consistent procedure for evaluating the statistical significance of the spectral
estimates. A significance can be estimated with a permutation test (Good, 2000), 
which has two main advantages: 1) it is distribution-free, and 2) it is applicable with
any spectral estimator. The main disadvantage is that this method can be very computer-intensive, 
particularly with long time series. The implementation of the permutation test for spectral analysis
requires the following steps:
\begin{enumerate}
\item 
From the original sequence of data $\{x(n), n = 1, 2, \dots, N\}$, 
the power spectrum, $S(\nu_i)$, is computed for a set of frequencies $\{\nu_i\}$.
\item
A high number, $M$, of random sequences are obtained by the random permutation of
the original data sequence, $\{x(n)\}$. For this work we used $M = 50,000$.
Each sequence, $\{x_j(n)\}$, obtained by random permutation has exactly the same
data as the original sequence, but in a different order.
\item $M$ different power spectra, $S_j(\nu_i)$, are calculated from the $M$ permutation sequences.
The power spectrum is estimated using the same order $p$ and for the same set of frequencies,
$\{\nu_i\}$, as the original power spectrum.
\item The statistical significance of the spectral features of the original sequence is assessed 
for each frequency, $\nu_i$, by the achieved significance level (ASL) defined as 
ASL($\nu_i$) = $I\{S(\nu_i) \ge S_j(\nu_i)\}/M$, where $I\{A\}$ is the indicator function of event $A$
(Pardo-Ig\'uzquize \& Rodriguez-Tovar 2005).
$I\{A\}$ is equal to 1 if event $A$ happens and 0 otherwise. 
Spectral values $S(\nu_i)$ with ASL($\nu_i$) $< \alpha$ are considered statistically significant with
confidence 100(1 - $\alpha$)\%, where the significance level $\alpha$ is chosen to be 0.001 (99.9\% confidence
level) in this work.
\end{enumerate}

\begin{figure}
\plotone{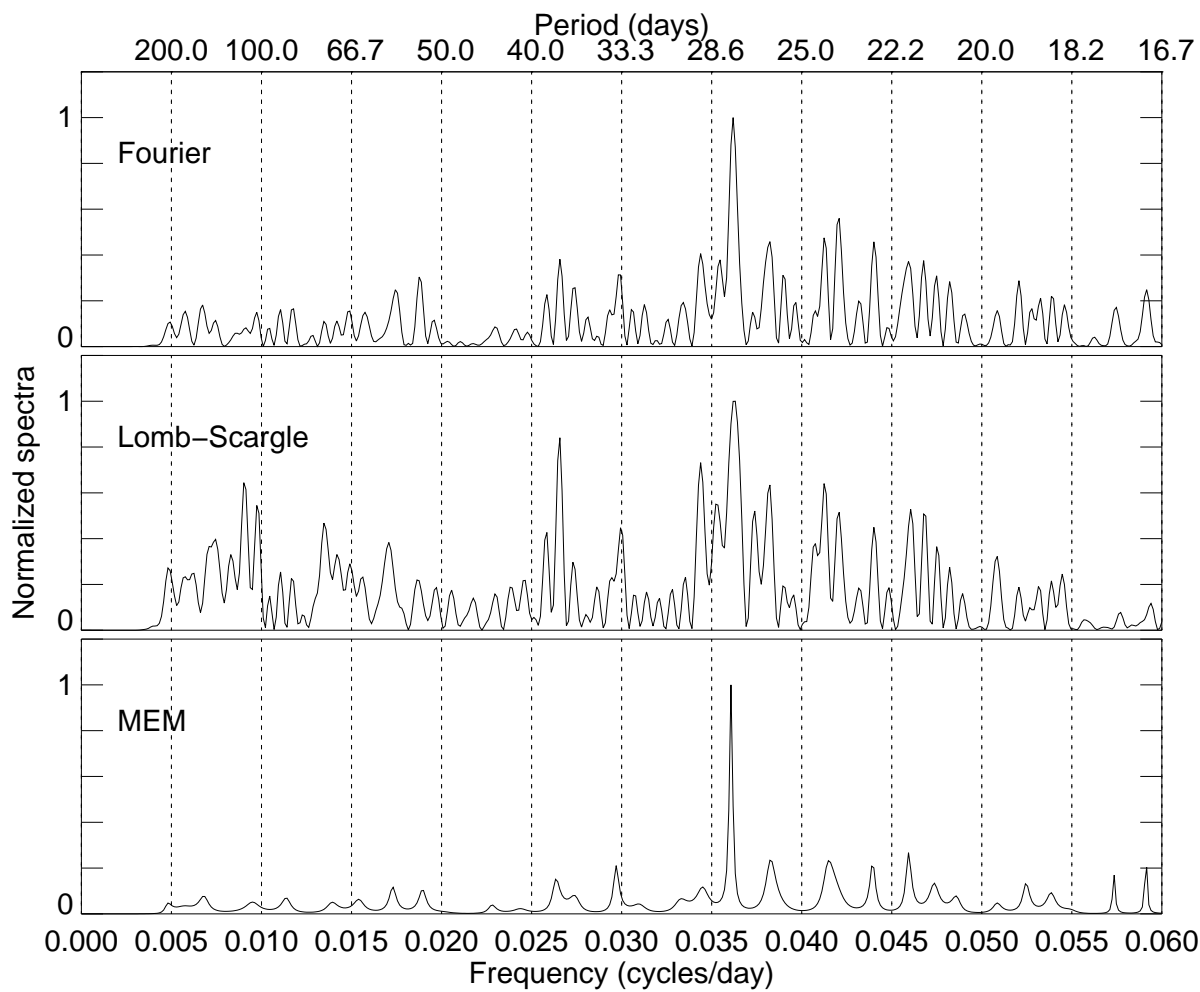}
\caption{Comparison of three different spectral estimators applied to the full-length 1-\AA~EM
time series. The same re-sampled and interpolated
data was used for both the Fourier and MEM analyzes, while for the Lomb-Scargle periodogram we
used the one-a-day measurements (the closest to Noon) and preserved the original time stamp. The frequencies at which
the power was computed are the same for all three estimations. Each power spectrum is normalized to its maximum
value.
\label{mem_others}}
\end{figure}

For comparison with the spectral estimation method described in this section, we computed
the power spectrum of the full-length 1-\AA~ EM time series using the direct Fourier transform and the
Lomb-Scargle periodogram. This comparison is shown in Figure \ref{mem_others}. The same re-sampled and interpolated
time series was used for the Fourier and MEM analyzes, while for the Lomb-Scargle periodogram we
used the one-a-day measurements (closest to Noon) and preserved the original time stamp. The frequencies at which
the power was computed are the same for all three estimations. All three methods detect similar spectral
features, but the MEM is able to isolate the $\sim$ 27-day rotational component from the time series 
with a much higher signal-to-noise ratio. For example, within the investigated spectral range, 
the ratio between the primary and secondary peaks is about a factor five for the MEM while it is
significantly lower for the other two estimators. In addition, the
Lomb-Scargle periodogram clearly shows a much more complex spectral structure compared to the other
two estimates.

\section{Results and Discussion}

The ISS parameter time series mostly cover
the extended minimum of Cycle 23/24 when the activity was low,
making the detection of SDR more difficult.
The main purpose of this study is to show that, under these less than favorable conditions, 
the selected technique of spectral analysis plays a substantial role in the ability to detect the 
signature of SDR. 
The major constraint imposed by the low level of solar activity
covered by this investigation is in the length of the temporal window that needs to be selected for the spectral
estimator. As discussed below, under these circumstances, at least 900 days of ISS data are required to provide
a (statistically) significant detection of SDR.
As the ISS program continues, the observations will cover the full duration of Cycle 24, and it is expected that shorter
time series will be needed. 

\subsection{Spectral Analysis}

We present here results from the spectral analysis performed on the full-length time series and
on the first and last 2/3 portions of the ISS data. To explore the time evolution of individual components
in the power spectrum we also subdivided every time series in overlapping segments, shifted by 1 day,
and computed the spectra of these segments (Section 4.2).
In addition, we applied 
the same analysis to a time series of disk-averaged longitudinal magnetic field flux density (MF) measurements derived
from daily SOLIS Vector Spectromagnetograph (VSM) magnetograms taken in the FeI 630.15 nm spectral line 
(Pietarila et al. 2012). 
The addition of an independent time series is necessary
to eliminate the possibility that our choice for the spectral estimator is biased towards a particular
data set. 
The selection of the MF time series was made because it provides the same time coverage
as the ISS observations, and also because the emission parameters derived from the Ca K spectral line 
are closely related to the unsigned magnetic flux (e.g., Ortiz \& Rast 2005).
The MF data are available from the SOLIS website at http://solis.nso.edu/vsm/vsm\_mnfield.html.
It is important to
emphasize here that the power spectra of individual time series are not expected to show the same
peaks at the same level of significance. There are two main reasons for that: 1) different parameters
are computed from different portions of the Ca II K line profile, and therefore have different
sensitivity to solar activity; 2) parameters computed from line-intensity values are more affected by possible
calibration issues than those computed from wavelength separations (see Table \ref{tbl} for a description
of the parameter used in this work). 

\begin{figure}
\plotone{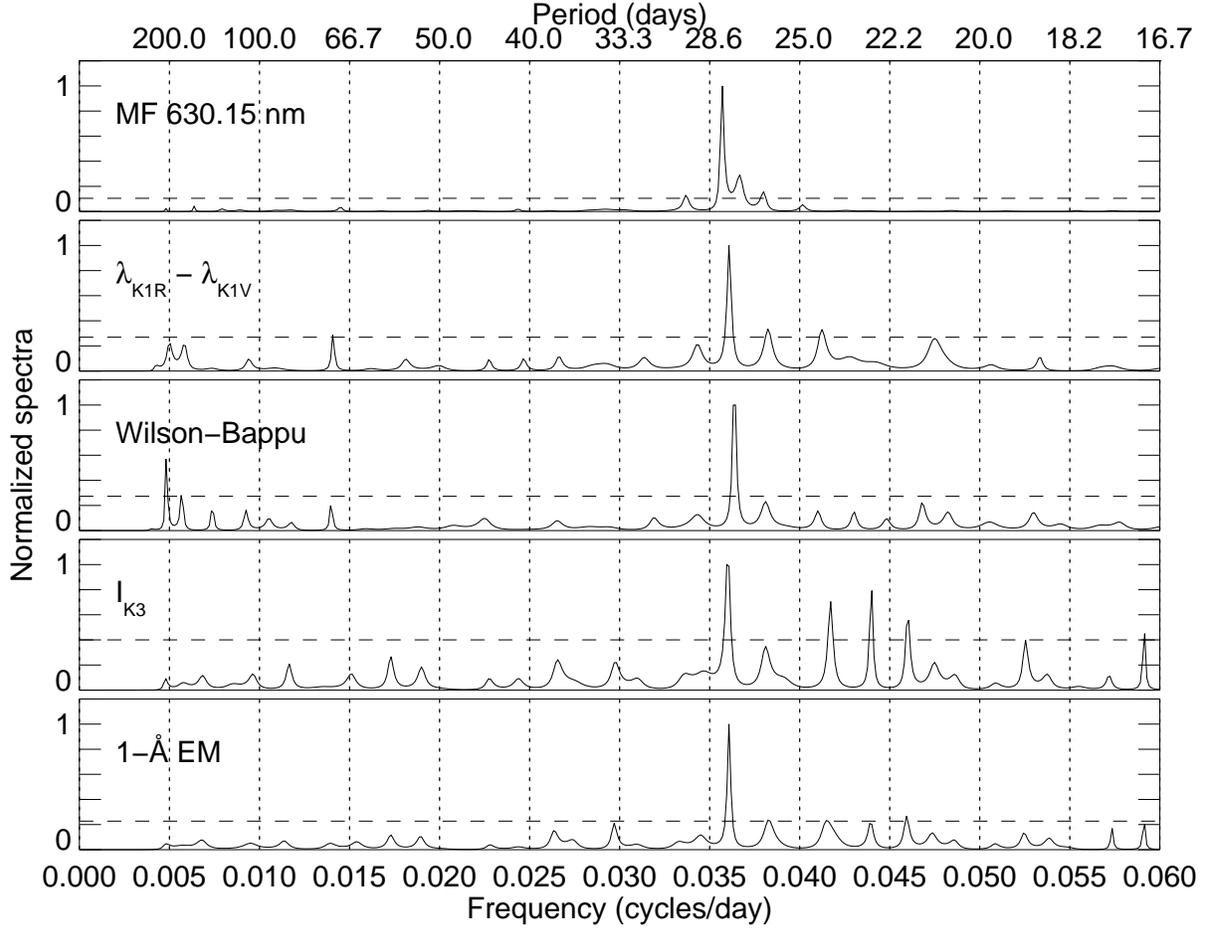}
\caption{Spectral estimation of the four full-length time series shown in Figure \ref{para_ts} and the
SOLIS-VSM FeI 630.15 nm mean longitudinal magnetic field flux density (MF). 
The horizontal
dashed-line in each spectrum indicates the 99.9\% confidence level. 
The list of rotation periods corresponding
to the tallest peak above the 99.9\% confidence level is given in Table \ref{tbl}, which also contains the results
from the other time series investigated in this study. 
\label{mem1_ts}}
\end{figure}

Figure \ref{mem1_ts} shows the power spectra of the full-length MF and four of the eight ISS full-length parameter time
series investigated in this study. All time series were divided by their standard deviation to get a zero
mean unit variance time series. 
The power spectrum of the MF time series is shown on the top, followed by
the two spectra corresponding to parameters computed from the wavelength separations of Ca K line features.
The bottom two spectra are Ca K intensity-related parameters. This Figure clearly shows the different features
associated to the two groups of parameters, and it is also representative for the other parameters listed in Table
\ref{tbl}. 
As shown in column 2 of Table \ref{tbl},
in eight of the nine time series investigated in this work the tallest peak above the 99.9\% confidence
level corresponds to a periodicity between 27.4 and 28.0 days. These numbers are within the range of values
expected for the solar rotation period. In general,
parameters defined in terms of wavelength separation typically show a spectrum with a 
very predominant peak, while intensity-related parameters show spectra with a more complex structure.
The only parameter showing a somewhat more erratic power spectrum is $\lambda_{\rm K2R} - \lambda_{\rm K2V}$,
the wavelength separation of the two emission maxima. The power spectrum
of this time series has a predominant peak corresponding to a rotational period of 141 days, most
likely the result of low sensitivity of the parameter to solar activity.

The detection of solar rotational periods in the power spectra of Ca II K parameter time series
can be affected by unrelated signatures of other processes: sunspot evolution (typically weeks),
faculae evolution (up to $\sim$60 days), and active longitudes characterized by the frequent emergence
of new magnetic flux (up to $\sim$200 days).
These processes can reduce the coherence of the rotational signal to the point where it is no longer
detectable, and some time series can be affected more than others. We believe this to be the case for the
power spectrum of $\lambda_{\rm K2R} - \lambda_{\rm K2V}$. In fact, a recent analysis (Pevtsov \& Bertello 2012) of 
basal and active sun emission components of the CaK line profiles indicates that the wavelength separation between 
the active Sun components does not change with the phase of the solar cycle. 

\begin{figure}
\plotone{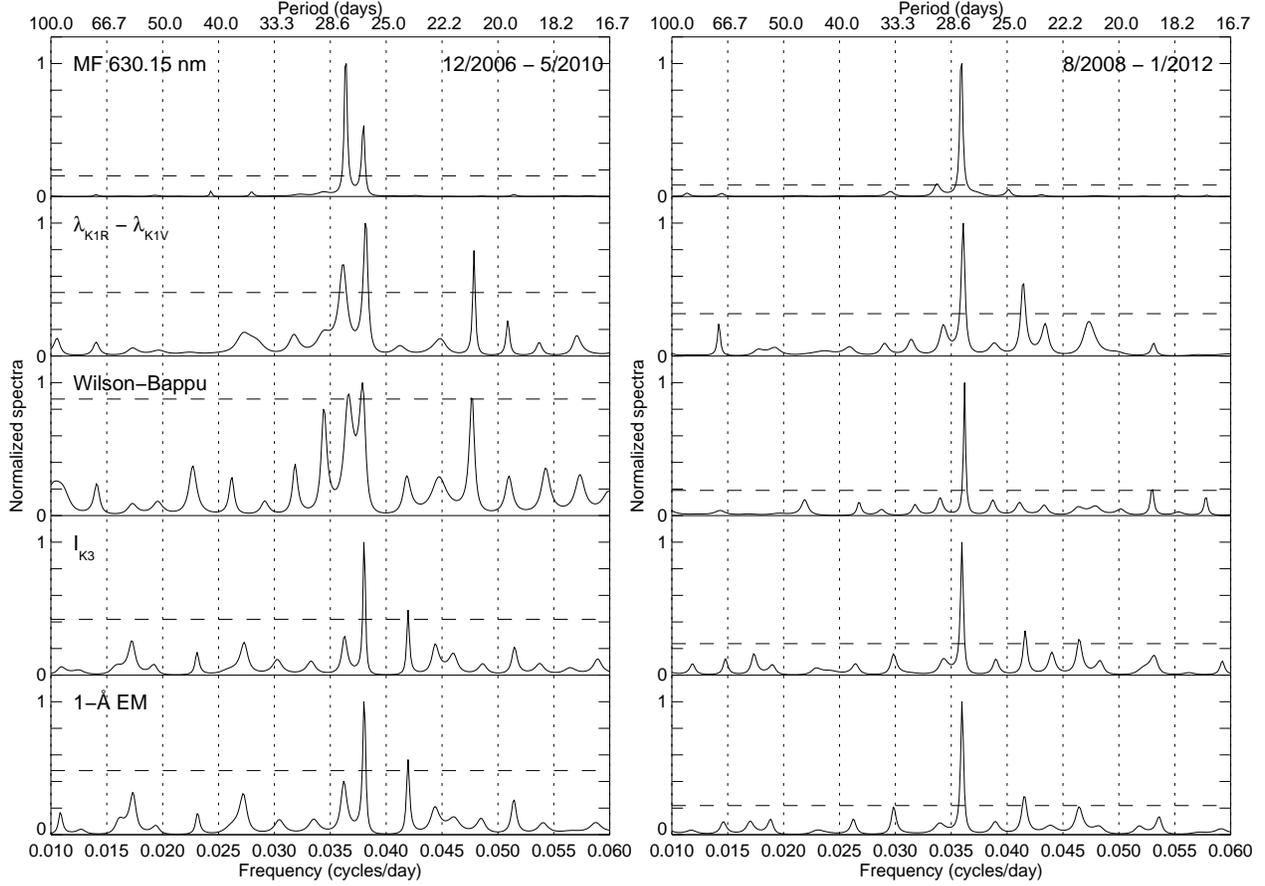}
\caption{Same as in Figure \ref{mem1_ts}, but the power spectra were
computed using the first (left) and the last 2/3 (right) portions 
of the full-length time series. The time interval covered by each segment is indicated in the plots.
The horizontal dashed-line indicates the 99.9\% confidence level. 
The list of rotation periods corresponding
to the tallest peak above the 99.9\% confidence level is given in Table \ref{tbl}, which also contains the results
from the other time series investigated in this study. 
\label{mem2_ts}}
\end{figure}

\begin{figure}
\plotone{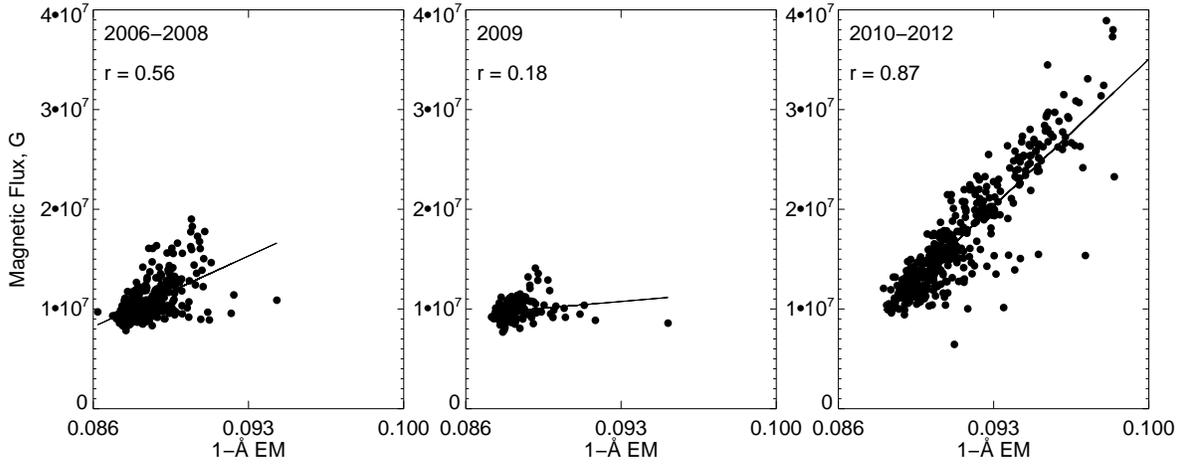}
\caption{Comparison of VSM total unsigned flux and ISS 1-\AA~ EM index for the
the declining phase of Cycle 23 (left), solar minimum (center), and raising phase of Cycle 24 (right).
Shown in solid line is the computed linear fit.
The Pearson correlation coefficient, r, is given in the upper left corner of each panel. 
We computed the $t$-value for the coefficient, r, to assess the statistical
significance of each correlation. The results are (for the left to the right panels): 
13.6 (2.59), 2.48 (2.69), and 37.3 
(2.59). Indicated in parenthesis are the expected 99\% cutoff values in the corresponding two-tailed
Student's $t$ distribution.
\label{vsm_iss}}
\end{figure}

We test the ability of our method to detect 
SDR during the solar cycle by computing the power spectra
of the first and last 2/3 portions of the individual time series. The results are shown in Figure \ref{mem2_ts}.
In the left column are the spectra computed from the first 2/3 portion of the data, while in the right column
we show the spectra from the last 2/3 portion of the time series. The second
portion of the data produces spectra characterized by the presence of a single predominant peak, while
the spectra of the first 3.5 years of data show a more complex structure.
This is not surprising since the first 2/3 of the data mainly cover the extended minimum of Cycle 23,
and therefore the signature of rotation is not as dominant. A comparison between the two
sets of spectra shows also a clear difference in the location of the tallest peak: around
$\sim$26.3 days for the first 3.5 years of data, consistent with activity near the equator
($\sim 7^{\circ}$, using the formula in Snodgrass \& Ulrich 1990), and $\sim$27.7 days
for the portion of the time series after August 2008. This distinct period of $\sim$27.7 days is in
agreement with features rotating at higher
latitude bands ($\sim 30^{\circ}$), as expected from the transition between Cycles 23 and 24.
The power spectrum of the first portion of the MF 630.15 nm time series shows two very
predominant peaks, located at 27.4 days and 26.3 days. While the secondary peak at 26.3 days
is in very good agreement with the results from the ISS time series for the same time frame,
the tallest peak at 27.4 days is more consistent with the ISS results from the second part of the data.
A possible interpretation is that late in the time series
covered by the first segment of the data, a new-cycle of activity was already present at higher latitudes. 
These high latitude fields were relatively weak, and did not 
contribute significantly to the disk-integrated Ca K profiles. Figure \ref{vsm_iss}
shows the relationship between the VSM total unsigned flux and the 1-\AA~ EM~ index for three different
periods. The correlation during the raising phase of Cycle 24 (2010-2012) is strong and statistically significant, while
during the declining phase of Cycle 23 (2007-2008) the correlation is weak, although still statistically significant.
During solar minimum (2009) no correlation is found. Therefore, the Ca K emission is a good proxy for magnetic activity
only for periods of intermediate and high levels of solar activity, which indirectly supports our above interpretation
that the presence of weak magnetic fields at high latitudes may explain the difference in periodicity
between the Ca K and MF results. 

\subsection{Time-frequency Analysis}

To further confirm the results shown in Figure \ref{mem2_ts}, we have performed a time-frequency
analysis to investigate the time evolution of the individual components of the power spectra
computed from the different time series. We have subdivided every time series in overlapping segments
of 900 days, with a 1-day time shift between consecutive segments, and computed the power spectra
according to the recipe described in the previous section. According to our tests with temporal windows
of different sizes, a
900-day sliding window provides the best compromise between time and frequency resolution. Figures
\ref{stack_k3}-\ref{stack_vsm} show time-frequency plots computed for two of the ISS parameter
time series and for the MF data. These plots are representative of the other time series.
The signature of SDR is clearly visible in these plots: most of the contribution to the power spectral
density of the signal is concentrated into a narrow frequency band, whose central value varies in 
time in the range 0.035 to 0.038 cycles/day (period range from 28.6 to 26.3 days). This change in period over
time is consistent with the migration of activity from latitudes bands at $\sim$ 35 degrees to the equator.
To large degree,
this curve is remarkably continuous and smooth at all times.
These
figures also show, for the ISS measurements, some additional power distribution over different time-frequency 
regions. This is likely
the result of some of the physical processes, e.g., sunspot and facular evolution, 
discussed in the next section, and possible
artifacts related to data calibration.
A close look at the time-frequency distributions around the period of solar rotation
shows some differences in the temporal behavior of the
power spectra. For example, the rotational period during
the declining phase of Cycle 23 deduced from the MF data seems to be slightly longer, and
more persistent, than
the one derived from the ISS measurements. This could be explained by changes in the correlation between magnetic
field and Ca K proxies with activity level (Figure \ref{vsm_iss}).

\subsection{Uncertainty in the Rotational Period}

The detection of SDR from disk-integrated measurements is affected not only by the
physical processes mentioned above but also by the particular analysis performed on the data.
With the approach described in this paper, the main factors limiting the identification
and accuracy of the rotational components in the power spectrum are: (1) the resampling and gap filling
of the signal,
(2) the choice of the order $p$ of the AR process used to model the time series, and (3) the length of the
time series window. The effect of the resampling and gap filling has been tested by computing the
power spectrum of the window function. We did not find any significant contribution from the window
function to the power spectral density of the signals in the region of interest, between $\sim$ 22 and
$\sim$ 33 days.
The order $p$ mainly changes the width of the rotational peaks (higher values of $p$ produce
narrower rotational peaks) without changing (significantly) their central frequency.
We can determine
the uncertainty in the deduced rotational period from the width of the peaks in the power spectrum and from the
variation in their central location as a result of small changes (10\%) in the selected order $p$ of the
AR process. Using this approach, the estimated uncertainty of the values listed in Table 1 is
approximately $\pm$0.3 days. 

\subsection{Length of the Time Series}

The choice for the length of the time series window used in the analysis described in section 3 
depends on three major factors: 1) the nature and properties of the data, 2) the level of activity,
and 3) the period of rotation. For the case of the Sun, a slowly rotating star, our results using ISS data show that
a 900-day window was necessary to provide a statistically significant detection of SDR over the 
whole period covered by the observations, which coincide with a prolong period of extremely low activity. 
Shorter time series (few hundred days) may
affect the statistical significance of the peaks and impact the overall
spectral distribution by producing spectra with a distribution of peaks over a broad spectral range. Furthermore,
the peaks may
often be inconsistent from one observing window to another, making it very
difficult to disentangle contributions from different sources. Using chromospheric full-disk Ca II K SPO
data in 200-day intervals, Donahue \& Keil (1995) were only able to detect rotation in 42\% of the observing
windows, mostly during periods of high solar activity. 

To investigate the effect of the activity level on the detectability of solar rotation on
shorter time intervals, we tested both the Lomb-Scargle periodogram
and the MEM on a sequence of 250-day ISS data segments. The MEM performed significantly better than 
the Lomb-Scargle periodogram, producing spectra of the same quality as those shown on the right panel of Figure
\ref{mem2_ts} with a clear detection of the rotational peak for about 38\% of the observing windows during periods
of both medium and high solar activity. The determined rotation periods from these spectra are consistent 
with the results obtained using a 900-day sliding time window. This test shows that the MEM can effectively
be used with relatively short time series, provided there is enough activity for the signature of
rotation to be present in the signal.
  
The application of this method to stellar measurements is more complex, but we see no reasons this
technique would not work effectively also for the stellar case. Quasi-continuous Ca II K and H
stellar measurements from ground-based observatories are typically limited in duration to a maximum of 
about 6 months. Our choice of 900-day and 250-day windows was mainly constrained by the fact that the Sun is 
a slowly rotating star and the ISS observations covered the extended solar minimum, with periods of no plages
or other weak remnants of active regions on the Sun. 

Young solar-like stars, however, are fast rotators with periods 
of a few days. Thus, for example, for a star with 5 days rotation period and low spot-activity 
it should require about (900/27)$\cdot$5 = 167 days
of observing time with a 58\% duty cycle. For a star with high spot-activity this number would be
(250/27)$\cdot$5 = 46 days. The above estimates refer to a detection of the rotation with a high significant level
of confidence; the rotation modulation can be detected with even shorter observing windows albeit at a lower
confidence level. This scenario
is slightly different for the case of stellar light curve time series.  
The rotational modulation in light curves is caused by dark spots as well
as bright faculae. In slowly rotating stars with relatively low level of activity, such as the Sun, 
these two contributions merge together to reduce the sensitivity of the signal to rotation modulation. 
On the other hand,
dark spots dominate active and young stars while faculae are predominant in weakly active stars during activity
minimum. In these two cases the signature of rotation can be observed from light curves if a sufficiently high
signal-to-noise ratio is achieved. 
Several ground- and spaced-based projects provide light curves over months or years that can be investigated
using the proposed technique. In particular, with a lifetime of at least 3.5 years and nearly continuous observations,
the Kepler mission (Borucki et al. 2010) is delivering stellar light curves of unprecedented time sampling
and photometric precision.

\subsection{Simplified Flux Transport Model}

To further validate our selection of parameters for spectral analysis, and to investigate the effects of 
properties of active Sun features on the detectability of SDR, 
we synthesized artificial data using a simplified flux transport model. 
The model evolves the radial magnetic field by the effects of flux emergence, 
differential rotation, meridional flow, and diffusion.
First, we created a “magnetogram” representing the entire solar surface covering 360 degrees 
in longitude and 180 degrees in latitude. The magnetogram is made of two components 
representing strong and weak magnetic fields. For weak fields we use a random distribution 
of small (in flux and size) bipolar elements.  
To simulate stronger fields of active regions, we insert larger bipolar pairs in selected latitudes and longitudes.  
The magnetogram is evolved by applying the known sunspot profile of SDR (Newton \& Nunn 1951), diffusion, 
and meridional circulation. Throughout the simulations, the magnetogram is “updated” 
by inserting new active regions. 
The location (latitude and longitude) of simulated active regions, 
their size and rate of emergence are taken from actual sunspot observations 
taken at the
Kislovodsk Solar Station of Pulkovo Observatory in Russia (http://www.solarstation.ru/?lang=en\&id=lastdata)
during the same period as the
ISS observations analyzed in this study (December 2, 2006-January 27, 2012). 
The magnetic flux of simulated active regions is defined using the known relation (Houtgast \& van Sluiters 1948) 
between sunspot area (S, in millionth of solar hemisphere) and magnetic flux ($\Phi$ in Gauss): 
$$\Phi = {{3700\times S} \over {S + 66}}.$$
Since the Kislovodsk dataset does not contain information about the tilt or polarity orientation of active regions, 
for the modeling we assume that all low latitude ($\leq$ 20 degrees) active regions observed prior 
to year 2009 (approximate time of solar minimum) are oriented in agreement with the Hale polarity 
rule for solar Cycle 23. High latitude active regions observed prior to 2009 and all regions 
observed from 2009-2012 are assigned a polarity orientation in correspondence with the solar Cycle 24. 
For simplicity, all active regions are assigned the same tilt relative to the equator. 
We experimented with tilts following Joy’s law (Hale et al. 1919) but found no significant
differences in the results.  
New active regions are "inserted" in the simulated magnetogram based on the time of the first observation
of the region. 
As an additional simplification, emergence of multiple active regions on the same day was 
ignored (only one active region was inserted in this limited number of cases). 
Unlike in more realistic flux transport models (e.g., Jiang et al. 2011), diffusion in our model is done 
by convolving the “magnetogram” with an inverse Gaussian filter. 

To investigate the role of diffusion in deriving solar rotation from Sun-as-a-star data, 
we run the model with six spatial filters corresponding to different parametrizations for diffusion (width
d of Gaussian filter): d=5, 10, 15, 30, 40, and 50.  
We found that d=30 results in the complete disappearance 
of an active region with strong magnetic flux over approximately 5- 6 solar rotations, 
thus providing a proxy for a reasonable solar diffusion rate. Based on this, 
we refer to models run with d=5, 10, 15 as cases of "fast"
diffusion, and d=40, 50 as cases of "slow" diffusion. As a qualitative reference, 
with d=5 newly injected active region magnetic field completely disappears in 2-3 days, 
and with d=50 the presence of diffusing flux is seen for 10+ solar rotations. 

Using the set of simulated 360 degree $\times$ 180 degree maps, we create full disk line-of-sight magnetograms  
for each day of observations during the ISS observing period. 
These daily magnetograms are used to create a time series of Sun-as-a-star unsigned and signed fluxes. 
Each series covers 1883 days (starting on December 2, 2006 and ending on February 27, 2012) 
and has a uniform time cadence of one day. 
The simulated time sequences are re-sampled as the ISS observations and subjected to the same data analysis.

Examples of time-frequency plots computed for the cases of "slow" (d = 40) and "fast" (d = 5) diffusion are shown
in Figures \ref{stack_1} and \ref{stack_2}. The case of "fast" diffusion has no clear
signature of solar rotation, indicating that the fast diffusing features do not live
long enough to be detected by our analysis. This case of "fast" diffusion may explain previous failed 
attempts in detecting differential rotation from disk integrated white-light spectra. 
Although the white-light observations have a contribution from features (such as sunspots) that follow solar rotation, 
these features decay rapidly (to small pore sizes) and their contribution to the white-light disk-integrated 
signal is diminished. 

When the diffusion time increases (Figure \ref{stack_2}), 
our technique is able to unambiguously detect the signature of differential rotation. The excellent agreement
between the results from the observed and simulated data is a further validation of the method described
in this paper. 
In addition, we also experimented with randomly injecting magnetic bipoles with a varying rate of emergence. 
We found that even when the active region emergence rate is low (up to one active region in 300+ days), 
we were able to detect the solar rotation for normal (d=30) and slow diffusion rates. 
For rapid diffusion (d=5), we found that the rotation rate is not well detectable even at very high 
rate of emergence of new active regions.
To investigate the effect of active longitudes (when active regions emerge preferentially at two 
longitudes separated by about 180 degrees), we run our code with randomly emerging active regions near 
these two preferred longitudes. 
Our analysis shows no definite splitting in rotation frequencies, which would indicate the presence of 
active longitudes.

\section{Conclusions} 

\begin{figure}
\plotone{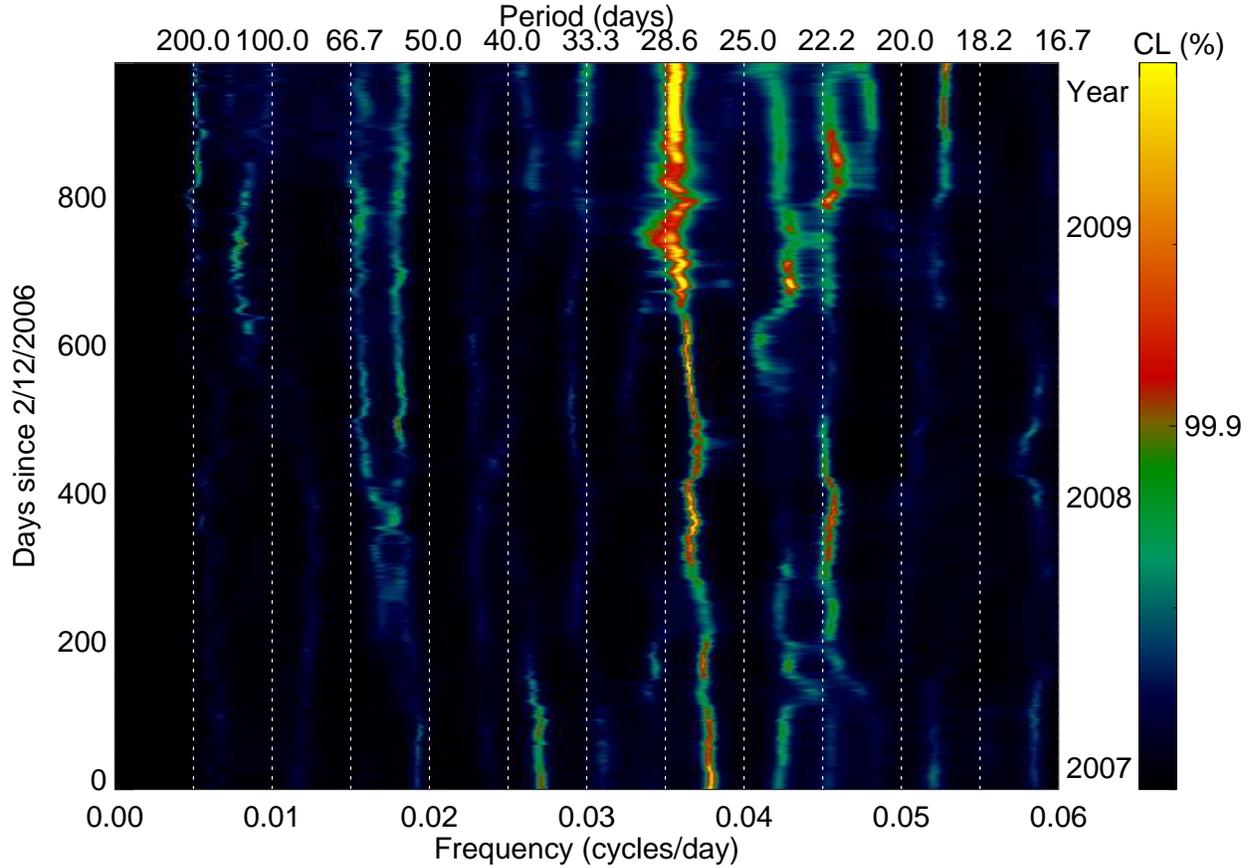}
\caption{ISS Ca II K3 core intensity time-frequency distribution from observations taken during the period 12/2/2006-1/27/2012. A 900-day sliding time window was used, with a difference between consecutive
segments of 1 day. The beginning of each segment, in days since 2/12/2006, 
is indicated on the left y-axis with the year shown on the right y-axis. 
The power spectral density above/below the 99.9\% confidence level (CL) is indicated
by the color bar. 
\label{stack_k3}}
\end{figure}

\begin{figure}
\plotone{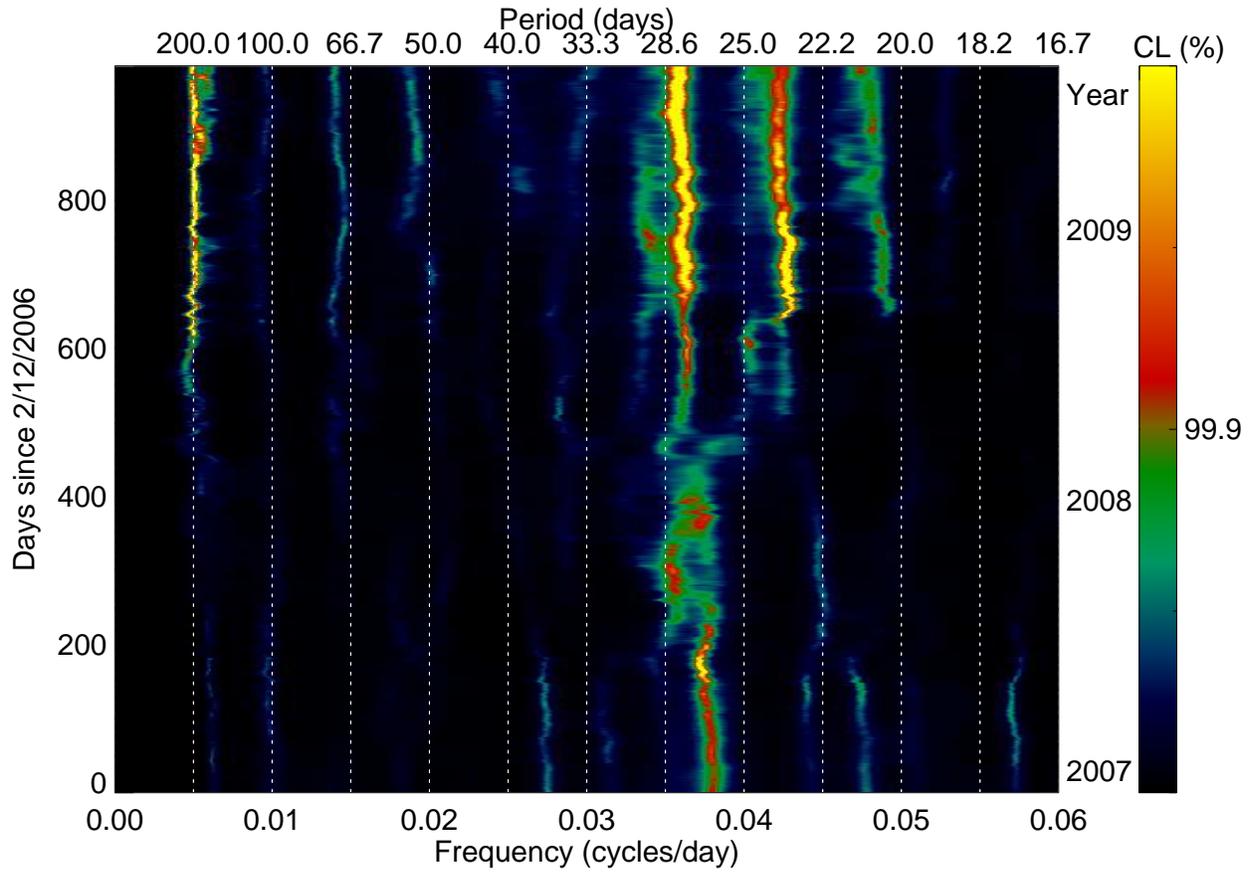}
\caption{Same as Figure \ref{stack_k3} but for the ISS $\lambda_{\rm K1R} - \lambda_{\rm K1V}$     
parameter.
\label{stack_k1dif}}
\end{figure}

\begin{figure}
\plotone{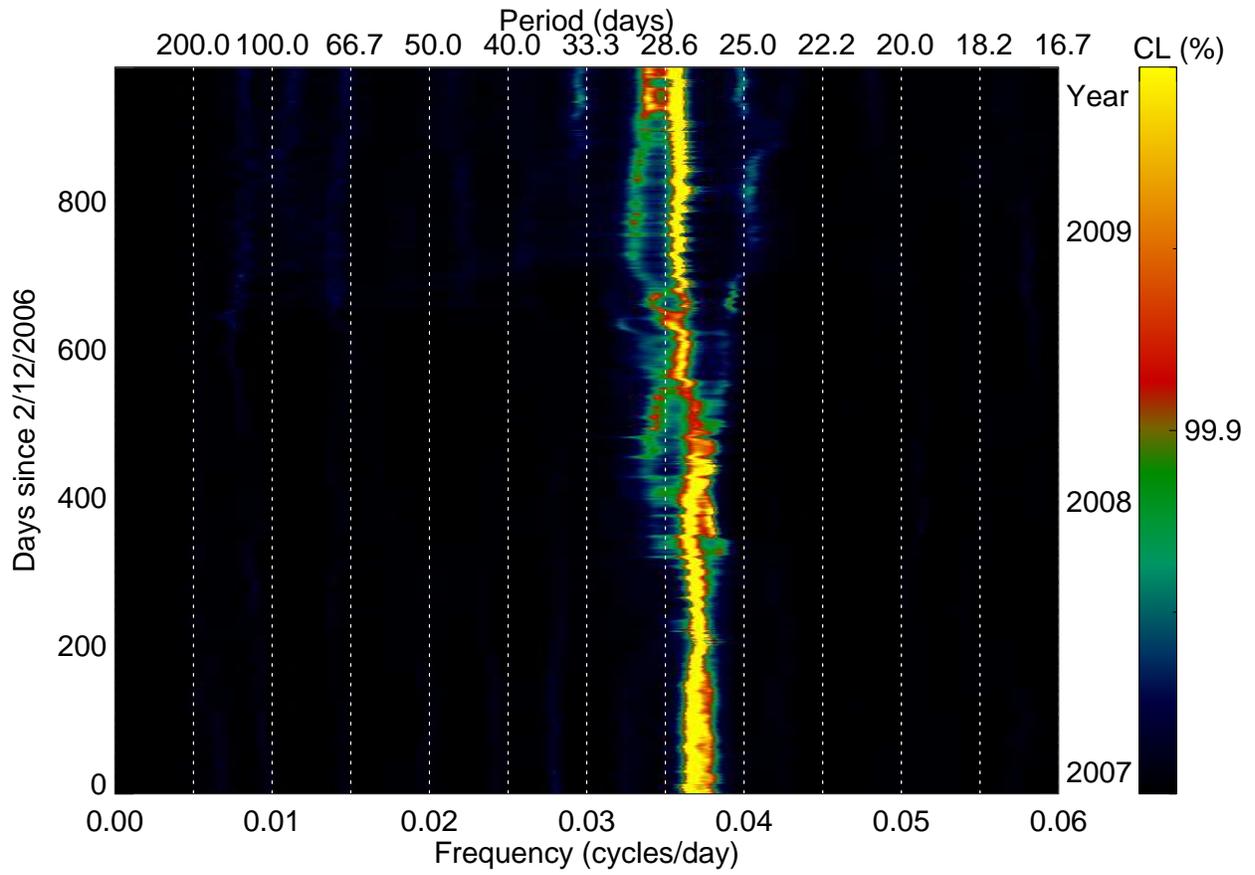}
\caption{Same as Figure \ref{stack_k3}, but for the 
SOLIS-VSM FeI 630.15 nm mean longitudinal magnetic field flux density time series (MF).
\label{stack_vsm}}
\end{figure}

\begin{figure}
\plotone{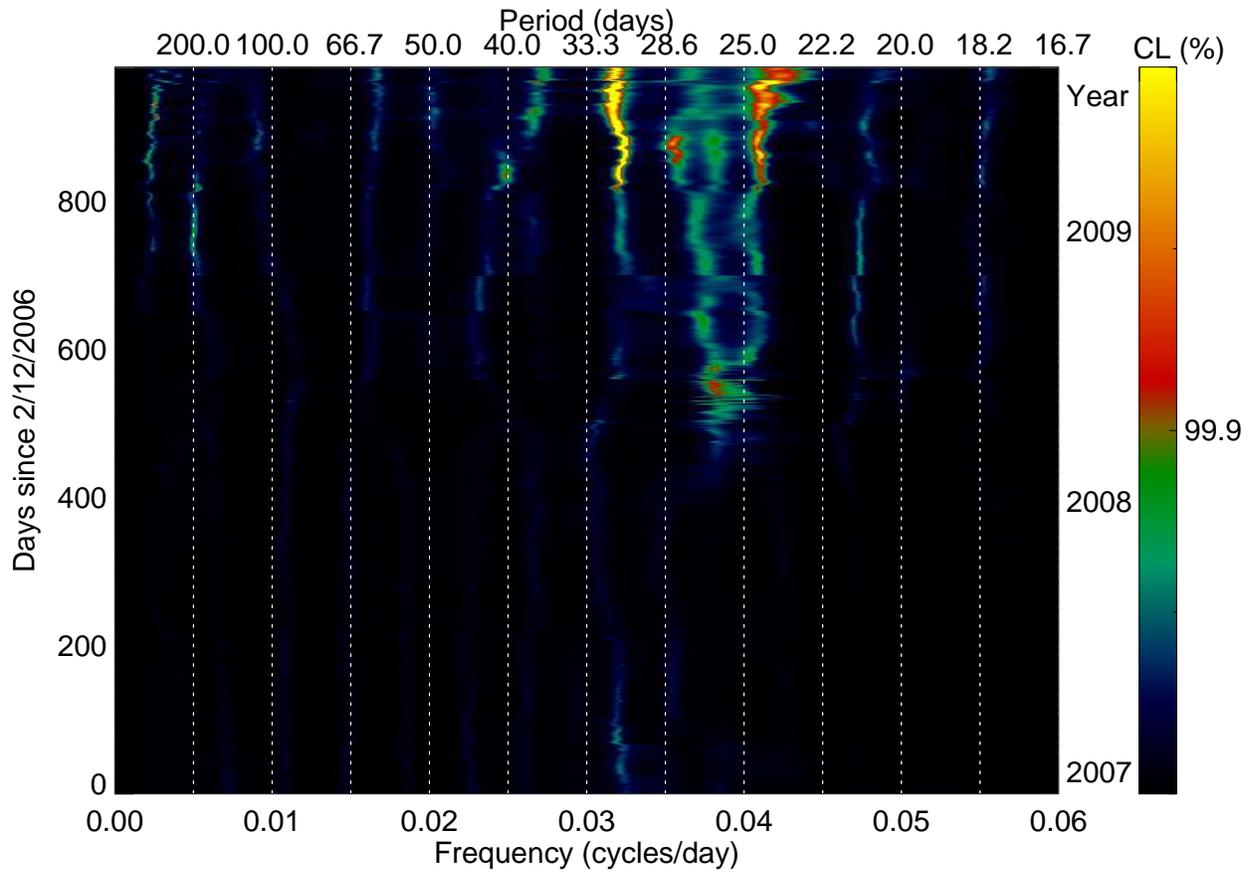}
\caption{Same as Figure \ref{stack_k3}, but for artificial data from a simplified flux transport model
with "fast" diffusion (d=5, see text).
\label{stack_1}}
\end{figure}

\begin{figure}
\plotone{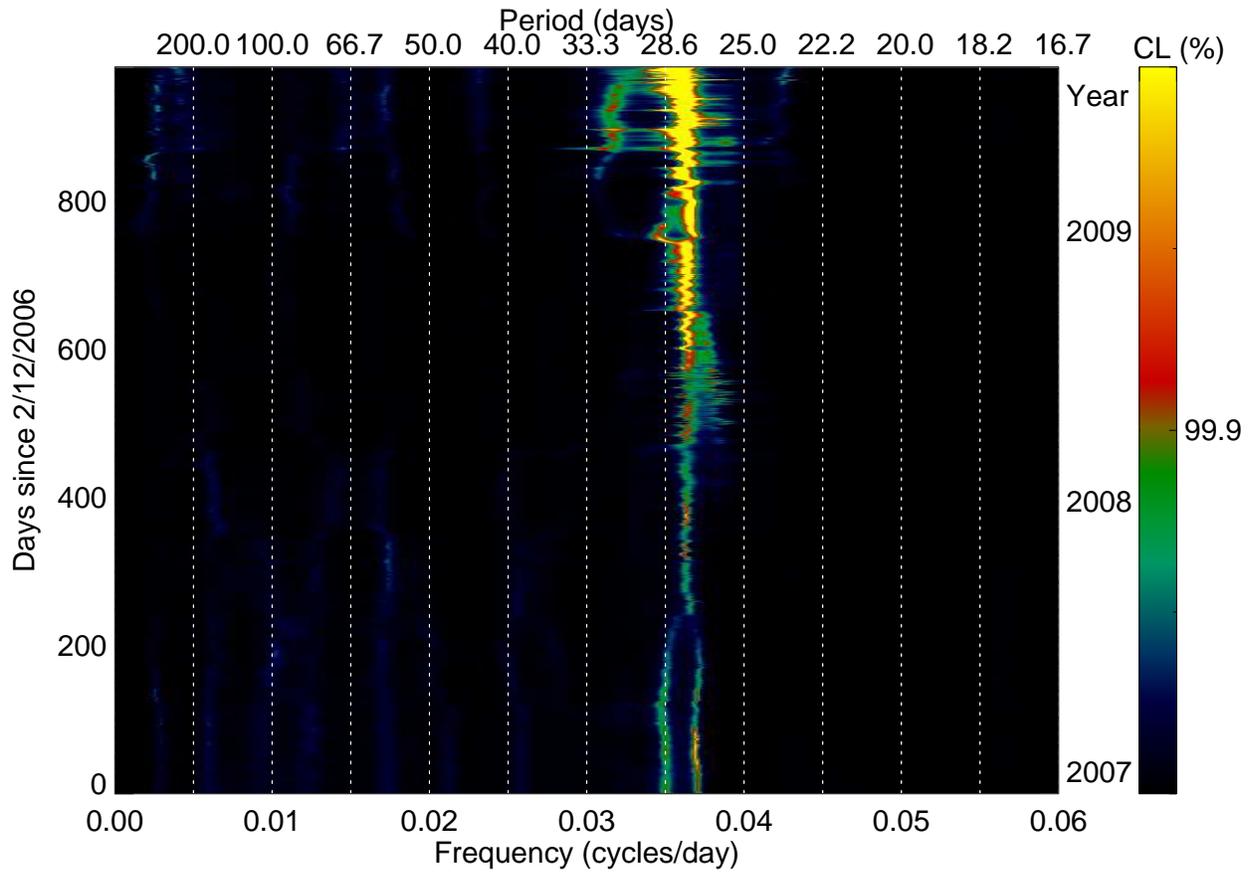}
\caption{Same as Figure \ref{stack_1}, but for the case of "slow" diffusion (d=40).
\label{stack_2}}
\end{figure}

In this paper we show that even under unfavorable conditions of very low levels of
solar activity it is possible to detect what can be characterized as the signature of surface differential
rotation from disk-integrated 
measurements. We present evidence of strong components in the power spectra computed from 
time series of parameters derived from
the temporal variations of Ca II K ISS line profiles and the MF time series that are consistent
with the rotation rate of active regions migrating on the solar disk
during the declining phase of Cycle 23 and
the rising phase of Cycle 24. The length of the time series window
used in our analysis is significantly longer than the one adopted in previous studies, but essential 
to characterize the pattern of differential rotation 
during periods of low magnetic activity. Longer time series also have the advantage of producing power spectra
with a reduced number of significant peaks over a large spectral band, making it easier to identify
the different contributions to the signal. In our tests, the adopted spectral estimator has been proven to be
significantly more effective than the Lomb-Scargle periodogram method in providing spectra with a reduced
number of significant peaks. This is very important when prior knowledge of the rotation rate profile
is uncertain, as is the case for stellar observations.
As some methods of spectral analysis may produce multiple peaks in a broad range of periods, it is important to correctly 
identify the peak corresponding to a true stellar rotation. 
The results from the ISS observations and from our numerical model suggest that, 
in addition to the significance level, 
the continuity of a specific frequency in power spectra can be used as a criterion. 
For example, Figure \ref{stack_k3} shows the presence of several peaks of different duration, but 
the peak around 27-28 days is the only one that persists throughout the entire data set.

The results of our numerical modeling suggest 
that the diffusion rate of active regions plays the most important role in detection of solar rotation
from Sun-as-a-star observations. 
Although we use a simplified flux transport model, our results show
the significance of the diffusion time-scale
parameter of active regions in the context of Sun-as-a-star differential rotation, with possible ramifications
to stellar applications as well. 
The rate of emergence and the presence of active longitudes seem to play a less relevant role.
Due to the simplifications in the adopted transport model, our results are mostly qualitative in nature. An important challenge therefore is to quantitatively assess the implications of diffusion
on determining SDR from disk-integrated measurements using more realistic flux transport models, 
a study that is beyond the scope of this paper.

Although some differences exist in the results obtained from the various time series investigated
in this study, the overall excellent agreement in the detection of SDR is a strong validation of
the adopted method. The use of simulated data, with a known rotation profile, is an important
confirmation of the quality of our methodology.
We believe that this approach can be successfully applied to other time series,
particularly stellar measurements, even when the duty-cycle is not very high. In fact, our re-sampled
ISS time series have a duty cycle of approximately 58\%.

\acknowledgments
SOLIS data used here are produced cooperatively by NSF/NSO and NASA/LWS.
The NSO and NOAO are operated by the Association of University for Research in Astronomy,
Inc. (AURA), under cooperative agreements with the National Science Foundation.

\clearpage

\begin{table}
\begin{center}
\caption{Measured synodic rotation periods calculated
from the power spectra of the eight ISS Ca II K parameter time series, and
SOLIS-VSM FeI 630.15 nm mean longitudinal magnetic field flux density (MF). 
Rotational periods have been calculated from spectra of the full-length time series (column 2),
and from spectra computed using the first (column 3) and the last (column 4) 2/3 portions of the
time series.
Only the period of the tallest peaks above the 99.9\% confidence level is listed. The estimated uncertainty of these
values is $\pm$0.3 days. \label{tbl}}
\begin{tabular}{lccc}
         &      \\
\tableline\tableline
Parameter         &  Full-length  & First 2/3  & Last 2/3 \\
                  & \multicolumn{3}{c}{Rotation period (days)} \\
\tableline
  1-\AA~EM        &   27.7  & 26.3 & 27.8 \\
   0.5-\AA~EM     &   27.7  & 26.3 & 27.7 \\
 I$_{\rm K3}$     &   27.8  & 26.3 & 27.8 \\
  Wilson-Bappu    &   27.4  & 26.4 & 27.6 \\
 $\lambda_{\rm K1R} - \lambda_{\rm K1V}$                    & 27.7 & 26.2 & 27.7 \\
 (I$_{\rm K2V}$-I$_{\rm K3}$)/(I$_{\rm K2R}$-I$_{\rm K3}$)  & 27.5 & 26.0 & 27.7 \\
 $\lambda_{\rm K2R} - \lambda_{\rm K2V}$                    & 141.0& none & none  \\
  I$_{\rm K2V}$/I$_{\rm K3}$                                & 27.7 & 51.3 & 27.8 \\
MF 630.15 nm                                                & 28.0 & 27.4 & 27.8 \\
\tableline
\end{tabular}
\tablecomments{1-\AA~EM is the 1-\AA~ emission index,
 0.5-\AA~EM  is the 0.5-\AA~ emission index,
 I$_{\rm K3}$ is the intensity in the line core,
 Wilson-Bappu is the wavelength separation between outer edges of K2,
 $\lambda_{\rm K1R} - \lambda_{\rm K1V}$ is the wavelength separation of V \& R emission minimum,
 (I$_{\rm K2V}$-I$_{\rm K3}$)/(I$_{\rm K2R}$-I$_{\rm K3}$) is the line asymmetry,
 $\lambda_{\rm K2R} - \lambda_{\rm K2V}$ is the wavelength separation of V \& R emission maximum, and
 I$_{\rm K2V}$/I$_{\rm K3}$ is the ratio of the blue emission peak to the intensity in the line core.
The violet and red wings of the line profile are indicated by V and R respectively.}
\end{center}
\end{table}

\clearpage

\clearpage

\clearpage

\clearpage

\clearpage

\clearpage

\clearpage

\clearpage

\clearpage

\clearpage

\begin{thebibliography}{40}
\expandafter\ifx\csname natexlab\endcsname\relax\def\natexlab#1{#1}\fi

\bibitem[{{Ammler-von Eiff} \& {Reiners}(2012)}]{2012A&A...542A.116A}
{Ammler-von Eiff}, M., \& {Reiners}, A. 2012, \aap, 542, A116

\bibitem[{{Balasubramaniam} \& {Pevtsov}(2011)}]{2011SPIE.8148E...8B}
{Balasubramaniam}, K.~S., \& {Pevtsov}, A. 2011, in Society of Photo-Optical
  Instrumentation Engineers (SPIE) Conference Series, Vol. 8148, Society of
  Photo-Optical Instrumentation Engineers (SPIE) Conference Series, 814809

\bibitem[{{Barnes} {et~al.}(2005){Barnes}, {Collier Cameron}, {Donati},
  {James}, {Marsden}, \& {Petit}}]{2005MNRAS.357L...1B}
{Barnes}, J.~R., {Collier Cameron}, A., {Donati}, J.-F., {James}, D.~J.,
  {Marsden}, S.~C., \& {Petit}, P. 2005, \mnras, 357, L1

\bibitem[{{Beck}(2000)}]{2000SoPh..191...47B}
{Beck}, J.~G. 2000, \solphys, 191, 47

\bibitem[{{Benedict} {et~al.}(2000){Benedict}, {Nobach}, \&
  {Tropea}}]{Benedict2000}
{Benedict}, L.~H., {Nobach}, H., \& {Tropea}, C. 2000, {\it Meas. Sci.
  Technol.}, 11, no. 8, 1089

\bibitem[{{Bertello} {et~al.}(2011){Bertello}, {Pevtsov}, {Harvey}, \&
  {Toussaint}}]{2011SoPh..272..229B}
{Bertello}, L., {Pevtsov}, A.~A., {Harvey}, J.~W., \& {Toussaint}, R.~M. 2011,
  \solphys, 272, 229

\bibitem[{{Borucki} {et~al.}(2010){Borucki}, {Koch}, {Basri}, {Batalha},
  {Brown}, {Caldwell}, {Caldwell}, {Christensen-Dalsgaard}, {Cochran},
  {DeVore}, {Dunham}, {Dupree}, {Gautier}, {Geary}, {Gilliland}, {Gould},
  {Howell}, {Jenkins}, {Kondo}, {Latham}, {Marcy}, {Meibom}, {Kjeldsen},
  {Lissauer}, {Monet}, {Morrison}, {Sasselov}, {Tarter}, {Boss}, {Brownlee},
  {Owen}, {Buzasi}, {Charbonneau}, {Doyle}, {Fortney}, {Ford}, {Holman},
  {Seager}, {Steffen}, {Welsh}, {Rowe}, {Anderson}, {Buchhave}, {Ciardi},
  {Walkowicz}, {Sherry}, {Horch}, {Isaacson}, {Everett}, {Fischer}, {Torres},
  {Johnson}, {Endl}, {MacQueen}, {Bryson}, {Dotson}, {Haas}, {Kolodziejczak},
  {Van Cleve}, {Chandrasekaran}, {Twicken}, {Quintana}, {Clarke}, {Allen},
  {Li}, {Wu}, {Tenenbaum}, {Verner}, {Bruhweiler}, {Barnes}, \&
  {Prsa}}]{2010Sci...327..977B}
{Borucki}, W.~J., {et~al.} 2010, Science, 327, 977

\bibitem[{{Brockwell} \& {Davis}(1991)}]{1991Brock}
{Brockwell}, P.~J., \& {Davis}, R.~A. 1991, Times Series: Theory and Methods,
  2nd edn. (Springer, New York)

\bibitem[{{Broersen}(2009)}]{2009Broersen}
{Broersen}, P.~M.~T. 2009, {\it IEEE Trans. Instrum. Meas.}, 58, 1380

\bibitem[{{de Waele} \& {Broersen}(2000)}]{2000Waele}
{de Waele}, S., \& {Broersen}, P.~M.~T. 2000, {\it IEEE Trans. Instrum. Meas.},
  49, 216

\bibitem[{{Donahue} \& {Keil}(1995)}]{1995SoPh..159...53D}
{Donahue}, R.~A., \& {Keil}, S.~L. 1995, \solphys, 159, 53

\bibitem[{{Donahue} {et~al.}(1996){Donahue}, {Saar}, \&
  {Baliunas}}]{1996ApJ...466..384D}
{Donahue}, R.~A., {Saar}, S.~H., \& {Baliunas}, S.~L. 1996, \apj, 466, 384

\bibitem[{{Frasca} {et~al.}(2011){Frasca}, {Fr{\"o}hlich}, {Bonanno},
  {Catanzaro}, {Biazzo}, \& {Molenda-{\.Z}akowicz}}]{2011A&A...532A..81F}
{Frasca}, A., {Fr{\"o}hlich}, H.-E., {Bonanno}, A., {Catanzaro}, G., {Biazzo},
  K., \& {Molenda-{\.Z}akowicz}, J. 2011, \aap, 532, A81

\bibitem[{{Fr{\"o}hlich} {et~al.}(2012){Fr{\"o}hlich}, {Frasca}, {Catanzaro},
  {Bonanno}, {Corsaro}, {Molenda-{\.Z}akowicz}, {Klutsch}, \&
  {Montes}}]{2012arXiv1205.5721F}
{Fr{\"o}hlich}, H.-E., {Frasca}, A., {Catanzaro}, G., {Bonanno}, A., {Corsaro},
  E., {Molenda-{\.Z}akowicz}, J., {Klutsch}, A., \& {Montes}, D. 2012, ArXiv
  e-prints

\bibitem[{{Good}(2000)}]{2000Good}
{Good}, P. 2000, Permutation Test: A Practical Guide to Resampling Methods for
  Testing Hypotheses, 2nd edn. (Springer, New York)

\bibitem[{{Hale} {et~al.}(1919){Hale}, {Ellerman}, {Nicholson}, \&
  {Joy}}]{1919ApJ....49..153H}
{Hale}, G.~E., {Ellerman}, F., {Nicholson}, S.~B., \& {Joy}, A.~H. 1919, \apj,
  49, 153

\bibitem[{{Hasler} {et~al.}(2002){Hasler}, {R{\"u}diger}, \&
  {Staude}}]{2002AN....323..123H}
{Hasler}, K.-H., {R{\"u}diger}, G., \& {Staude}, J. 2002, Astronomische
  Nachrichten, 323, 123

\bibitem[{{Hempelmann}(2002)}]{2002A&A...388..540H}
{Hempelmann}, A. 2002, \aap, 388, 540

\bibitem[{{Hempelmann} \& {Donahue}(1997)}]{1997A&A...322..835H}
{Hempelmann}, A., \& {Donahue}, R.~A. 1997, \aap, 322, 835

\bibitem[{{Houtgast} \& {van Sluiters}(1948)}]{1948BAN....10..325H}
{Houtgast}, J., \& {van Sluiters}, A. 1948, \bain, 10, 325

\bibitem[{{Jiang} {et~al.}(2011){Jiang}, {Cameron}, {Schmitt}, \&
  {Sch{\"u}ssler}}]{2011A&A...528A..83J}
{Jiang}, J., {Cameron}, R.~H., {Schmitt}, D., \& {Sch{\"u}ssler}, M. 2011,
  \aap, 528, A83

\bibitem[{{Keil} \& {Worden}(1984)}]{1984ApJ...276..766K}
{Keil}, S.~L., \& {Worden}, S.~P. 1984, \apj, 276, 766

\bibitem[{{Koch} {et~al.}(2010){Koch}, {Borucki}, {Basri}, {Batalha}, {Brown},
  {Caldwell}, {Christensen-Dalsgaard}, {Cochran}, {DeVore}, {Dunham},
  {Gautier}, {Geary}, {Gilliland}, {Gould}, {Jenkins}, {Kondo}, {Latham},
  {Lissauer}, {Marcy}, {Monet}, {Sasselov}, {Boss}, {Brownlee}, {Caldwell},
  {Dupree}, {Howell}, {Kjeldsen}, {Meibom}, {Morrison}, {Owen}, {Reitsema},
  {Tarter}, {Bryson}, {Dotson}, {Gazis}, {Haas}, {Kolodziejczak}, {Rowe}, {Van
  Cleve}, {Allen}, {Chandrasekaran}, {Clarke}, {Li}, {Quintana}, {Tenenbaum},
  {Twicken}, \& {Wu}}]{2010ApJ...713L..79K}
{Koch}, D.~G., {et~al.} 2010, \apjl, 713, L79

\bibitem[{{Maeder} \& {Eenens}(2004)}]{2004IAUS..215.....M}
{Maeder}, A., \& {Eenens}, P., eds. 2004, IAU Symposium, Vol. 215, {Stellar
  Rotation}

\bibitem[{{McClellan} {et~al.}(1979){McClellan}, {Parks}, \&
  {Rabiner}}]{1979McClellan}
{McClellan}, J.~H., {Parks}, T., \& {Rabiner}, L. 1979, In: Programs for
  Digital Signal Processing, IEEE Press, 5.1

\bibitem[{{Newton} \& {Nunn}(1951)}]{1951MNRAS.111..413N}
{Newton}, H.~W., \& {Nunn}, M.~L. 1951, \mnras, 111, 413

\bibitem[{{Ortiz} \& {Rast}(2005)}]{2005MmSAI..76.1018O}
{Ortiz}, A., \& {Rast}, M. 2005, \memsai, 76, 1018

\bibitem[{{Papoulis}(1991)}]{1991Papoulis}
{Papoulis}, A. 1991, Probability, Random Variables and Stochastic Processes,
  3rd edn. (McGraw-Hill Inc.)

\bibitem[{{Percival} \& {Walden}(1993)}]{1993Percival}
{Percival}, D.~B., \& {Walden}, A.~T. 1993, Spectral Analysis for Physical
  Applications. Multitaper and Conventional Univariate Techniques (Cambridge
  University Press, London)

\bibitem[{{Pevtsov} \& {Bertello}(2012)}]{2012AAS...22020309P}
{Pevtsov}, A.~A., \& {Bertello}, L. 2012, in American Astronomical Society
  Meeting Abstracts, Vol. 220, American Astronomical Society Meeting Abstracts,
  \#203.09

\bibitem[{{Pietarila} {et~al.}(2012){Pietarila}, {Bertello}, {Harvey}, \&
  {Pevtsov}}]{Pieta2012}
{Pietarila}, A., {Bertello}, L., {Harvey}, J., \& {Pevtsov}, A. 2012, \solphys,
  in press

\bibitem[{Robinson \& Treitel(2000)}]{Robin2000}
Robinson, E.~A., \& Treitel, S. 2000, Geophysical Signal Analysis (Society of
  Exploration Geophysicists)

\bibitem[{{Savanov} \& {Dmitrienko}(2012)}]{2012ARep...56..116S}
{Savanov}, I.~S., \& {Dmitrienko}, E.~S. 2012, Astronomy Reports, 56, 116

\bibitem[{{Scargle}(1982)}]{1982ApJ...263..835S}
{Scargle}, J.~D. 1982, \apj, 263, 835

\bibitem[{{Schrijver}(1996)}]{1996IAUS..176....1S}
{Schrijver}, C.~J. 1996, in IAU Symposium, Vol. 176, Stellar Surface Structure,
  ed. K.~G. {Strassmeier} \& J.~L. {Linsky}, 1

\bibitem[{{Schrijver} \& {Zwaan}(2000)}]{2000ssma.book.....S}
{Schrijver}, C.~J., \& {Zwaan}, C. 2000, {Solar and Stellar Magnetic Activity}
  (Cambridge University Press)

\bibitem[{{Snodgrass} \& {Ulrich}(1990)}]{1990ApJ...351..309S}
{Snodgrass}, H.~B., \& {Ulrich}, R.~K. 1990, \apj, 351, 309

\bibitem[{{Thompson} {et~al.}(2003){Thompson}, {Christensen-Dalsgaard},
  {Miesch}, \& {Toomre}}]{2003ARA&A..41..599T}
{Thompson}, M.~J., {Christensen-Dalsgaard}, J., {Miesch}, M.~S., \& {Toomre},
  J. 2003, \araa, 41, 599

\bibitem[{{Walkowicz} \& {Basri}(2011)}]{2011ASPC..448..177W}
{Walkowicz}, L.~M., \& {Basri}, G.~S. 2011, in Astronomical Society of the
  Pacific Conference Series, Vol. 448, Astronomical Society of the Pacific
  Conference Series, ed. C.~{Johns-Krull}, M.~K. {Browning}, \& A.~A. {West},
  177

\bibitem[{{Wallace} {et~al.}(2007){Wallace}, {Hinkle}, \&
  {Livingston}}]{2007assp.book.....W}
{Wallace}, L., {Hinkle}, K., \& {Livingston}, W. 2007, {An Atlas of the
  Spectrum of the Solar Photosphere from 13,500 to 33,980 cm$^{-1}$ (2942 to
  7405 {\AA})}

\end{thebibliography}
\end{document}